\documentclass[aps,prd,preprint,superscriptaddress,byrevtex]{revtex4}
\usepackage{graphicx} 
\usepackage{dcolumn}  
\usepackage{amsmath}
\usepackage{epstopdf}

\begin{document}


\preprint{\vbox{ \hbox{   }
                 \hbox{BELLE-CONF-0669}
}}

\title{ \quad \\[0.5cm] \boldmath Determination of $|V_{cb}|$ and
  $m_b$ from Inclusive\\
  $B\to X_c\ell\nu$ and $B\to X_s\gamma$~Decays at Belle}

\affiliation{Budker Institute of Nuclear Physics, Novosibirsk}
\affiliation{Chiba University, Chiba}
\affiliation{Chonnam National University, Kwangju}
\affiliation{University of Cincinnati, Cincinnati, Ohio 45221}
\affiliation{University of Frankfurt, Frankfurt}
\affiliation{The Graduate University for Advanced Studies, Hayama} 
\affiliation{Gyeongsang National University, Chinju}
\affiliation{University of Hawaii, Honolulu, Hawaii 96822}
\affiliation{High Energy Accelerator Research Organization (KEK), Tsukuba}
\affiliation{Hiroshima Institute of Technology, Hiroshima}
\affiliation{University of Illinois at Urbana-Champaign, Urbana, Illinois 61801}
\affiliation{Institute of High Energy Physics, Chinese Academy of Sciences, Beijing}
\affiliation{Institute of High Energy Physics, Vienna}
\affiliation{Institute of High Energy Physics, Protvino}
\affiliation{Institute for Theoretical and Experimental Physics, Moscow}
\affiliation{J. Stefan Institute, Ljubljana}
\affiliation{Kanagawa University, Yokohama}
\affiliation{Korea University, Seoul}
\affiliation{Kyoto University, Kyoto}
\affiliation{Kyungpook National University, Taegu}
\affiliation{Swiss Federal Institute of Technology of Lausanne, EPFL, Lausanne}
\affiliation{University of Ljubljana, Ljubljana}
\affiliation{University of Maribor, Maribor}
\affiliation{University of Melbourne, Victoria}
\affiliation{Nagoya University, Nagoya}
\affiliation{Nara Women's University, Nara}
\affiliation{National Central University, Chung-li}
\affiliation{National United University, Miao Li}
\affiliation{Department of Physics, National Taiwan University, Taipei}
\affiliation{H. Niewodniczanski Institute of Nuclear Physics, Krakow}
\affiliation{Nippon Dental University, Niigata}
\affiliation{Niigata University, Niigata}
\affiliation{University of Nova Gorica, Nova Gorica}
\affiliation{Osaka City University, Osaka}
\affiliation{Osaka University, Osaka}
\affiliation{Panjab University, Chandigarh}
\affiliation{Peking University, Beijing}
\affiliation{University of Pittsburgh, Pittsburgh, Pennsylvania 15260}
\affiliation{Princeton University, Princeton, New Jersey 08544}
\affiliation{RIKEN BNL Research Center, Upton, New York 11973}
\affiliation{Saga University, Saga}
\affiliation{University of Science and Technology of China, Hefei}
\affiliation{Seoul National University, Seoul}
\affiliation{Shinshu University, Nagano}
\affiliation{Sungkyunkwan University, Suwon}
\affiliation{University of Sydney, Sydney NSW}
\affiliation{Tata Institute of Fundamental Research, Bombay}
\affiliation{Toho University, Funabashi}
\affiliation{Tohoku Gakuin University, Tagajo}
\affiliation{Tohoku University, Sendai}
\affiliation{Department of Physics, University of Tokyo, Tokyo}
\affiliation{Tokyo Institute of Technology, Tokyo}
\affiliation{Tokyo Metropolitan University, Tokyo}
\affiliation{Tokyo University of Agriculture and Technology, Tokyo}
\affiliation{Toyama National College of Maritime Technology, Toyama}
\affiliation{University of Tsukuba, Tsukuba}
\affiliation{Virginia Polytechnic Institute and State University, Blacksburg, Virginia 24061}
\affiliation{Yonsei University, Seoul}
  \author{K.~Abe}\affiliation{High Energy Accelerator Research Organization (KEK), Tsukuba} 
  \author{K.~Abe}\affiliation{Tohoku Gakuin University, Tagajo} 
  \author{I.~Adachi}\affiliation{High Energy Accelerator Research Organization (KEK), Tsukuba} 
  \author{H.~Aihara}\affiliation{Department of Physics, University of Tokyo, Tokyo} 
  \author{D.~Anipko}\affiliation{Budker Institute of Nuclear Physics, Novosibirsk} 
  \author{K.~Aoki}\affiliation{Nagoya University, Nagoya} 
  \author{T.~Arakawa}\affiliation{Niigata University, Niigata} 
  \author{K.~Arinstein}\affiliation{Budker Institute of Nuclear Physics, Novosibirsk} 
  \author{Y.~Asano}\affiliation{University of Tsukuba, Tsukuba} 
  \author{T.~Aso}\affiliation{Toyama National College of Maritime Technology, Toyama} 
% ...
\collaboration{The Belle Collaboration}

\noaffiliation

\begin{abstract}
  We present an analysis of the Belle measured moments of the lepton
energy and hadronic mass spectra in $B\to X_c\ell\nu$~decays and the
photon energy spectrum in $B\to X_s\gamma$~decays using theoretical
expressions derived in the 1S and kinetic schemes. The magnitude of
the Cabibbo-Kobayashi-Maskawa matrix element $V_{cb}$, the $b$-quark
mass and other non-perturbative parameters are extracted. In the 1S
scheme analysis we find $|V_{cb}|=(41.49\pm 0.52{\rm (fit)}\pm
0.20(\tau_B))\times 10^{-3}$ and $m_b^\mathrm{1S}=(4.729\pm
0.048)$~GeV. In the kinetic scheme, we obtain $|V_{cb}|=(41.93\pm
0.65{\rm (fit)}\pm 0.07{\rm (\alpha_s)}\pm 0.63{\rm (th)})\times
10^{-3}$ and $m_b^\mathrm{kin}=(4.564\pm 0.076{\rm (fit)}\pm
0.003(\alpha_s))$~GeV.

\end{abstract}


\maketitle

\section{Introduction}

The most precise determinations of the Cabibbo-Kobayashi-Maskawa (CKM)
matrix element~$|V_{cb}|$~\cite{Kobayashi:1973fv} are obtained using
combined fits to inclusive $B$~decay
distributions~\cite{Bauer:2004ve,Buchmuller:2005zv}. These analyses
are based on calculations of the semileptonic decay rate in the
frameworks of the Operator Product Expansion (OPE)~\cite{wilson} and the Heavy Quark
Effective Theory (HQET)~\cite{Bauer:2004ve,Benson:2003kp} which
predict this quantity in terms of $|V_{cb}|$, the $b$-quark mass
$m_b$, and a few non-perturbative matrix elements that enter at the
order $1/m^2_b$.


Several studies have shown  that the spectator model decay rate, in which bound state effects are neglected, is the leading term in a well-defined 
expansion controlled by  the parameter $\Lambda _{\rm QCD}/m_b$ \cite{Benson:2003kp,gremm-kap,falk,Gambino:2004qm}. 
Non-perturbative corrections to this leading approximation arise only to order
 $1/m_b^2$. The key issue in this approach is the ability to separate non-perturbative 
 corrections (expressed as a series in powers of $1/m_b$), and perturbative 
 corrections (expressed in powers of  $\alpha _s$). 
There are various different methods 
to handle the energy scale $\mu$ used to separate long-distance from short-distance physics. 

The coefficients of the $1/m_b$ power terms are expectation values of operators that include non-perturbative physics.  In this framework, non-perturbative corrections are parameterised by quark masses and matrix elements of higher dimensional operators which are presently poorly known. 
The experimental accuracy already achieved, and that expected from larger data sets recorded by the $B$-factories, make the ensuing  theory uncertainty a major limiting factor. The extraction of the non-perturbative parameters describing the heavy 
quark masses, kinetic energy of the $b$ quark and the $1/m_b^3$ corrections directly from the data has therefore become a key issue. 

The non-calculable, non-perturbative quantities are parametrised in terms of expectation values of hadronic matrix elements, which can be related to the shape (characterised by moments) of inclusive decay spectra \cite{Bauer:2004ve,Gambino:2004qm,Benson:2004sg}. High precision comparison of theory and experiment requires a precise determination of the heavy quark masses, as well as the nonperturbative matrix elements that enter the expansion.  These are $\lambda_{1,2}$ or $\mu_{\pi,G}$ which parameterise the nonperturbative corrections to inclusive observables at $\mathcal{O}(\Lambda_{\rm{QCD}}^2/m_b^2)$.  At $\mathcal{O} (\Lambda_{\rm{QCD}}^3/m_b^3$), more matrix elements occur, denoted by $\rho_{1,2}$ and $\tau_{1-4}$ or $\rho_{D,LS}$.

In this paper we make use of the Heavy Quark Expansions (HQEs) that express the semileptonic decay width  $\Gamma_{\rm s.l.}$, moments of
the lepton energy and hadronic mass spectra in $B\to
X_c\ell\nu$~decays and the photon energy spectrum in $B\to
X_s\gamma$~decays in terms of the running kinetic quark masses $m_b^{\rm kin}$ and $m_c^{\rm kin}$ as well as the 1S $b$-quark mass $m_b^{\rm 1S}$.  Further details of these two schemes are discussed in later sections of the paper.  These schemes should ultimately yield consistent results for $|V_{cb}|$.  The precision of the $b$-quark mass is also important for the determination of $|V_{ub}|$, the least well understood element in the CKM matrix, and a limiting factor in the uncertainty on the unitarity triangle.


\section{Experimental Input}

Belle has measured the partial branching fractions~$\mathcal{B}(B\to
X_c\ell\nu)_{E_\ell>E_\mathrm{min}}$ and the first, second, third and
fourth moments of the truncated electron energy spectrum in $B\to
X_ce\nu$, $\langle E_\ell\rangle_{E_\ell>E_\mathrm{min}}$, $\langle
(E_\ell-\langle E_\ell\rangle)^2\rangle_{E_\ell>E_\mathrm{min}}$,
$\langle (E_\ell-\langle
E_\ell\rangle)^3\rangle_{E_\ell>E_\mathrm{min}}$ and $\langle
(E_\ell-\langle E_\ell\rangle)^4\rangle_{E_\ell>E_\mathrm{min}}$, for
nine different electron energy thresholds ($E_\mathrm{min}=0.4$, 0.6,
0.8, 1.0, 1.2, 1.4, 1.6, 1.8 and 2.0~GeV)~\cite{el}.

We have measured the first, second central and second non-central
moments of the hadron invariant mass squared ($M^2_X$)~spectrum in $B\to X_c\ell\nu$, $\langle
M^2_X\rangle_{E_\ell>E_\mathrm{min}}$, $\langle(M^2_X-\langle
M^2_X\rangle)^2\rangle_{E_\ell>E_\mathrm{min}}$ and $\langle
M^4_X\rangle_{E_\ell>E_\mathrm{min}}$ for seven different lepton
energy thresholds ($E_\mathrm{min}=0.7$, 0.9, 1.1, 1.3, 1.5, 1.7 and
1.9~GeV)~\cite{mx}.

For $B\to X_s\gamma$ we have measured the first and second moments of
the truncated photon energy spectrum , $\langle
E_\gamma\rangle_{E_\gamma>E_\mathrm{min}}$ and
$\langle(E_\gamma-\langle
E_\gamma\rangle)^2\rangle_{E_\gamma>E_\mathrm{min}}$. These
measurements are available for six minimum photon energies
($E_\mathrm{min}=1.8$, 1.9, 2.0, 2.1, 2.2 and
2.3~GeV)~\cite{Abe:2005cv}.

Hence, there are a total of 71 Belle
measurements of inclusive spectra available for use in the
fits~\cite{ref:1}. The measurements used in the 1S and kinetic
scheme fit analyses are shown in Table~\ref{tab:1}. We have excluded
measurements that do not have corresponding theoretical
predictions. Measurements with higher cutoff energies ({\it i.e.} electron energy and hadron mass moments with
$E_\mathrm{min}>1.5$~GeV and  photon energy moments with
$E_\mathrm{min}>2$~GeV) are not used to determine the HQE parameters, as theoretical predictions are not considered
reliable in this region. Finally, we have also excluded points where
correlations with neighbouring points are too high.
\begin{table}
  \begin{center}
    {\small \begin{tabular}{c|c|c}
      \hline \hline
      \rule[-1.3ex]{0pt}{4ex} & 1S scheme & kinetic scheme\\
      \hline
      \rule{0pt}{2.7ex} & $n=0$ $E_\mathrm{min}=0.6$, 1.0, 1.4 & $n=0$
      $E_\mathrm{min}=0.4$, 0.8\\
      Lepton moments~$\langle E^n_\ell\rangle_{E_\mathrm{min}}$ &
      $n=1$ $E_\mathrm{min}=0.6$, 0.8, 1.0, 1.2, 1.4 & $n=1$
      $E_\mathrm{min}=0.4$, 0.8, 1.0, 1.2 1.4\\
      & $n=2$ $E_\mathrm{min}=0.6$, 1.0, 1.4 & $n=2$
      $E_\mathrm{min}=0.4$, 0.8, 1.0, 1.2 1.4\\
      \rule[-1.3ex]{0pt}{1.3ex} & $n=3$ $E_\mathrm{min}=0.8$, 1.2 &
      $n=3$ $E_\mathrm{min}=0.4$, 0.8, 1.0, 1.2 1.4\\
      \hline
      \rule{0pt}{2.7ex}Hadron moments~$\langle
      M^{2n}_X\rangle_{E_\mathrm{min}}$ & $n=1$ $E_\mathrm{min}=0.7$,
      1.1, 1.3, 1.5 & $n=1$ $E_\mathrm{min}=0.7$, 0.9, 1.1, 1.3\\
      \rule[-1.3ex]{0pt}{1.3ex} & $n=2$ $E_\mathrm{min}=0.7$, 0.9,
      1.3 & $n=2$ $E_\mathrm{min}=0.7$, 0.9, 1.1, 1.3\\
      \hline
      \rule{0pt}{2.7ex}Photon moments~$\langle
      E^n_\gamma\rangle_{E_\mathrm{min}}$ & $n=1$ $E_\mathrm{min}=1.8$,
      2.0 & $n=1$ $E_\mathrm{min}=1.8$, 1.9, 2.0\\
      \rule[-1.3ex]{0pt}{1.3ex} & $n=2$ $E_\mathrm{min}=1.8$, 2.0
      & $n=2$ $E_\mathrm{min}=1.8$, 1.9, 2.0\\
      \hline \hline
    \end{tabular} }
  \end{center}
  \caption{Experimental input used in the 1S and kinetic scheme
    analyses. The values of $E_\mathrm{min}$ are given in GeV. The 1S
    (kinetic) scheme analysis uses a total of 24 (31) measurements.}
    \label{tab:1}
\end{table}

The value of $|V_{cb}|$ is dependent on the $B$~meson lifetimes. The
measured semileptonic ratios can be written as
$\mathcal{B}_\mathrm{s.l.}=\tau_\mathrm{eff}\Gamma_\mathrm{s.l.}$ in
terms of an effective lifetime,
$\tau_\mathrm{eff}=f_+\tau_++f_0\tau_0$. Using the most recent world
average values for the lifetimes and the $b$-hadron fractions, we
obtain $\tau_\mathrm{eff}=(1.585 \pm 0.006$)~ps~\cite{unknown:2006bi}.  For simplicity we refer to this quantity simply as $\tau_B$.

\section{1S Scheme Analysis}

\subsection{Theoretical input}

The inclusive spectral moments of $B\to X_c\ell\nu$ decays have been
derived in the 1S scheme up to
$\mathcal{O}(1/m_b^3)$~\cite{Bauer:2004ve}. The theoretical
expressions for the truncated moments have the following form (where
$\langle X\rangle_{E_\mathrm{min}}$ represents any of the experimental
observables):
\begin{equation}
  \begin{split}
    \langle X\rangle_{E_\mathrm{min}} =
    X^{(1)}+X^{(2)}\Lambda+X^{(3)}\Lambda^2+X^{(4)}\Lambda^3+X^{(5)}\lambda_1+X^{(6)}\Lambda\lambda_1+X^{(7)}\lambda_2+X^{(8)}\Lambda\lambda_2+X^{(9)}\rho_1\\+X^{(10)}\rho_2+X^{(11)}\tau_1+X^{(12)}\tau_2+X^{(13)}\tau_3+X^{(14)}\tau_4+X^{(15)}\varepsilon+X^{(16)}\varepsilon^2_{\rm{BLM}}+X^{(17)}\varepsilon\Lambda~.
  \end{split} \label{eq:1}
\end{equation}
The coefficients $X^{(k)}$, determined by theory, are functions of
$E_\mathrm{min}$. The non-perturbative corrections are parametrized by
$\Lambda$ ($\mathcal{O}(m_b)$), $\lambda_1$ and $\lambda_2$
($\mathcal{O}(1/m_b^2)$), and $\tau_1$, $\tau_2$, $\tau_3$, $\tau_4$,
$\rho_1$ and $\rho_2$ ($\mathcal{O}(1/m_b^3)$).

Predictions for the partial branching fractions are obtained using the
following expression,
\begin{equation}
  \mathcal{B}(B\to X_c\ell\nu)_{E_\mathrm{min}}=\langle
  X\rangle_{\mathcal{B},E_\mathrm{min}}\frac{\eta_{QED}|V_{cb}|^2G_F^2m^5}{192\pi^3\tau_B}~,
\end{equation}
where $\langle X\rangle_{\mathcal{B},E_\mathrm{min}}$ is an expression
of the form of Eq.~\ref{eq:1}, $m$ is the 1S reference mass,
$m=m_{\Upsilon(1S)}/2$, $\eta_{QED}=1.007$, and $G_F^2
m^5/(192\pi^3)=5.4\times 10^{-11}$.

In this analysis, we determine a total of seven parameters:
$|V_{cb}|$, $\Lambda$, $\lambda_1$, $\tau_1$, $\tau_2$, $\tau_3$ and
$\rho_1$. One of the higher order parameters, $\tau_4$ is set to zero
, and from available constraints, {\it e.g.} $B^*-B$~mass splitting, the remaining
parameters in Eq.~\ref{eq:1} are set to:
$\lambda_2=0.1227-0.0145\lambda_1$ and
$\rho_2=0.1361+\tau_2$, following advice from Ref.~\cite{Bauer:2004ve}. The parameter~$\Lambda$ is the difference
between the $b$-quark mass and the reference value about which it is
expanded, {\it i.e.}, $\Lambda=m_{\Upsilon(1S)}/2-m_b^{1S}$. We will
present our results in terms of $m_b^{1S}$ in place of $\Lambda$.

\subsection{\boldmath The $\chi^2$~function}

The fit takes into account both experimental and theoretical
uncertainties. Following the approach in Ref.~\cite{Bauer:2004ve}, an
element of the combined experimental and theoretical error matrix is
given by
\begin{equation}
  \sigma_{ij}^2=\sigma_i \sigma_j c_{ij},
\end{equation}
where $i$ and $j$ denote the observables and $c_{ij}$ is the
experimental correlation matrix element. The total error on the
observable~$i$ is defined as
\begin{eqnarray}
  \sigma_i = \sqrt{(\sigma_i^{\rm{exp}})^2 + (A f_n m_B^{2n})^2 +
    (B_i/2)^2}  \mbox{ for the $n$th hadron moment~,}\\
  \sigma_i = \sqrt{(\sigma_i^{\rm{exp}})^2 + (A f_n (m_B/2)^n)^2 +
    (B_i/2)^2}  \mbox{ for the $n$th lepton moment~,}\\
  \sigma_i = \sqrt{(\sigma_i^{\rm{exp}})^2 + (A f_n (m_B/2)^n)^2 +
    (B_i/2)^2}  \mbox{ for the $n$th photon moment~,}
\end{eqnarray}
and $f_0=f_1=1$, $f_2=1/4$ and $f_3=1/(6\sqrt{3})$. Here,
$\sigma_i^{\rm{exp}}$ are the experimental errors, $B_i=X^{(16)}$ are
the coefficients of the last computed terms in the perturbation series
(providing the error on the uncalculated higher order perturbative
terms). The dimensionless parameter $A$ contains various theoretical errors (uncalculated
power corrections, uncalculated effects of order
$(\alpha_s/4\pi)\Lambda_{\rm{QCD}}^2/m_b^2$, and effects not included
in the OPE, {\it i.e.}, duality violation), and is multiplied by dimensionful
quantities.  We fix $A=0.001$ as in Ref.~\cite{Bauer:2004ve}. For $B\to X_s\gamma$, the accessible phase space is
limited, and the theoretical extraction of $m_b$ is affected by shape
function effects. So, $A$ is multiplied by the ratio of the difference
from the end point relative to $E_\mathrm{min}=1.8$~GeV.

As the fit does not provide strong constraints on the
$\mathcal{O}(1/m_b^3)$~parameters, it is necessary to provide
constraints to ensure their convergence to sensible values. We achieve
this by introducing extra terms in the $\chi^2$~function of the fit,
\begin{equation}
  \chi^2_{\rm{param}}(m_\chi,M_\chi)=
  \begin{cases} 
    0 & |\mathcal{O}| \le m_\chi^3~,\\ 
    \left(|\langle\mathcal{O}\rangle|-m_\chi^3\right)^2/M_\chi^6 &
    |\mathcal{O}|>m_\chi^3~,
  \end{cases} 
\end{equation}
where ($m_\chi$,$M_\chi$) are both quantities of order
$\Lambda_{\rm{QCD}}$, and $\langle\mathcal{O}\rangle$ represents any
$\mathcal{O}(1/m_b^3)$~parameters. In the fit, we take
$M_\chi=m_\chi=500$~MeV after Ref.~\cite{Bauer:2004ve}.  The parameter $m_\chi$ may have a value anywhere
between 500~MeV and 1~GeV.

The overall form of the $\chi^2$~function used in the 1S~fit is
\begin{equation}
  \chi^2 = \sum_{i,j} (\langle X\rangle_i^\mathrm{meas.}-\langle
  X\rangle_i^\mathrm{1S})\mathrm{cov}^{-1}_{ij}(\langle
  X\rangle_j^\mathrm{meas.}-\langle X\rangle_j^\mathrm{1S})+
  \sum_{i=1}^2\chi^2_{\rho_i}+\sum_{i=1}^3\chi^2_{\tau_i}~,
  \label{eq:2}
\end{equation}
where $\langle X\rangle_i^\mathrm{meas.}$ are the measured moments and $\langle
  X\rangle_i^\mathrm{1S}$ are the corresponding 1S~scheme predictions.

\subsection{Fit results and discussion}

\begin{figure}
  \includegraphics[width=0.45\textwidth]{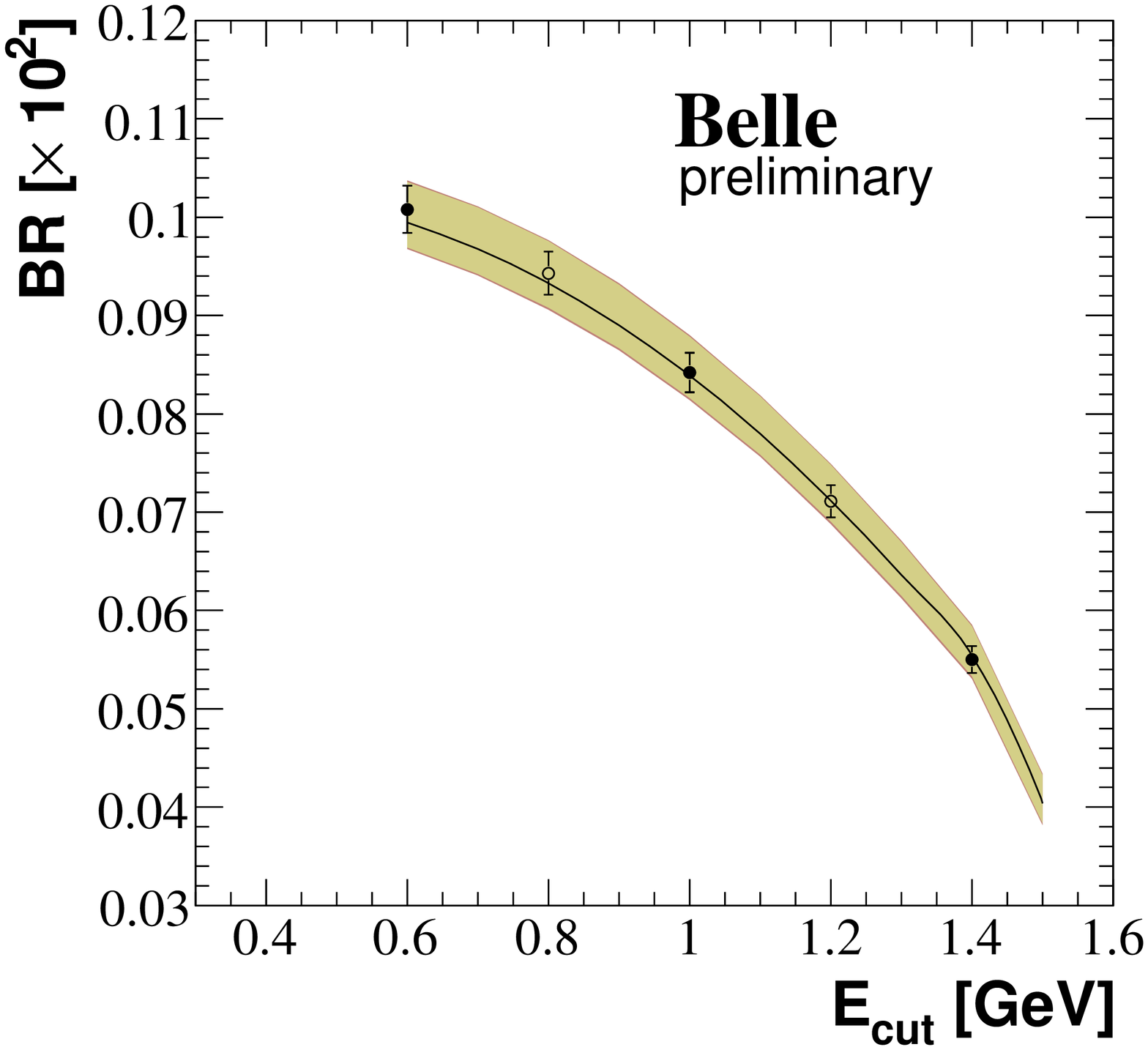}
  \includegraphics[width=0.45\textwidth]{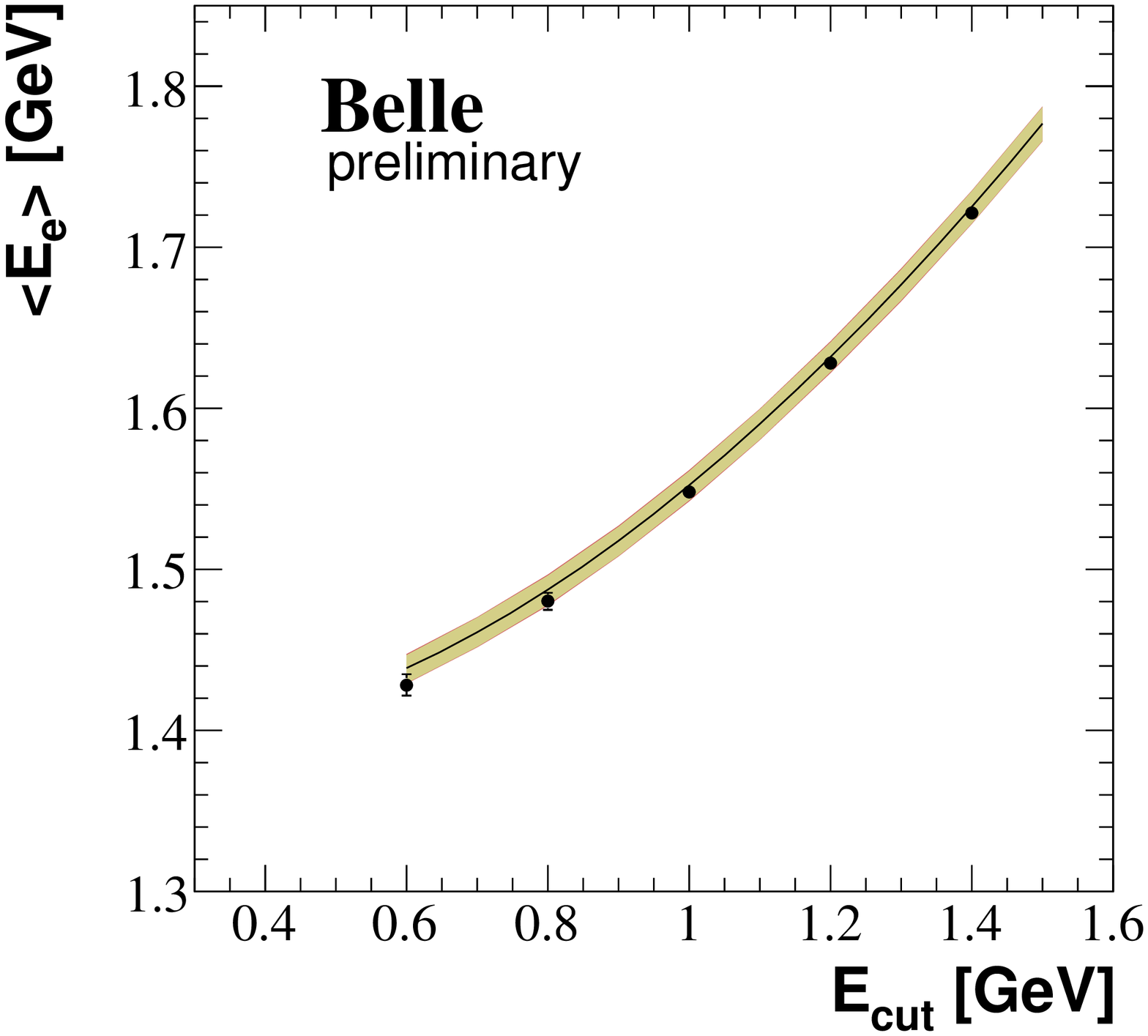}\\
  \includegraphics[width=0.45\textwidth]{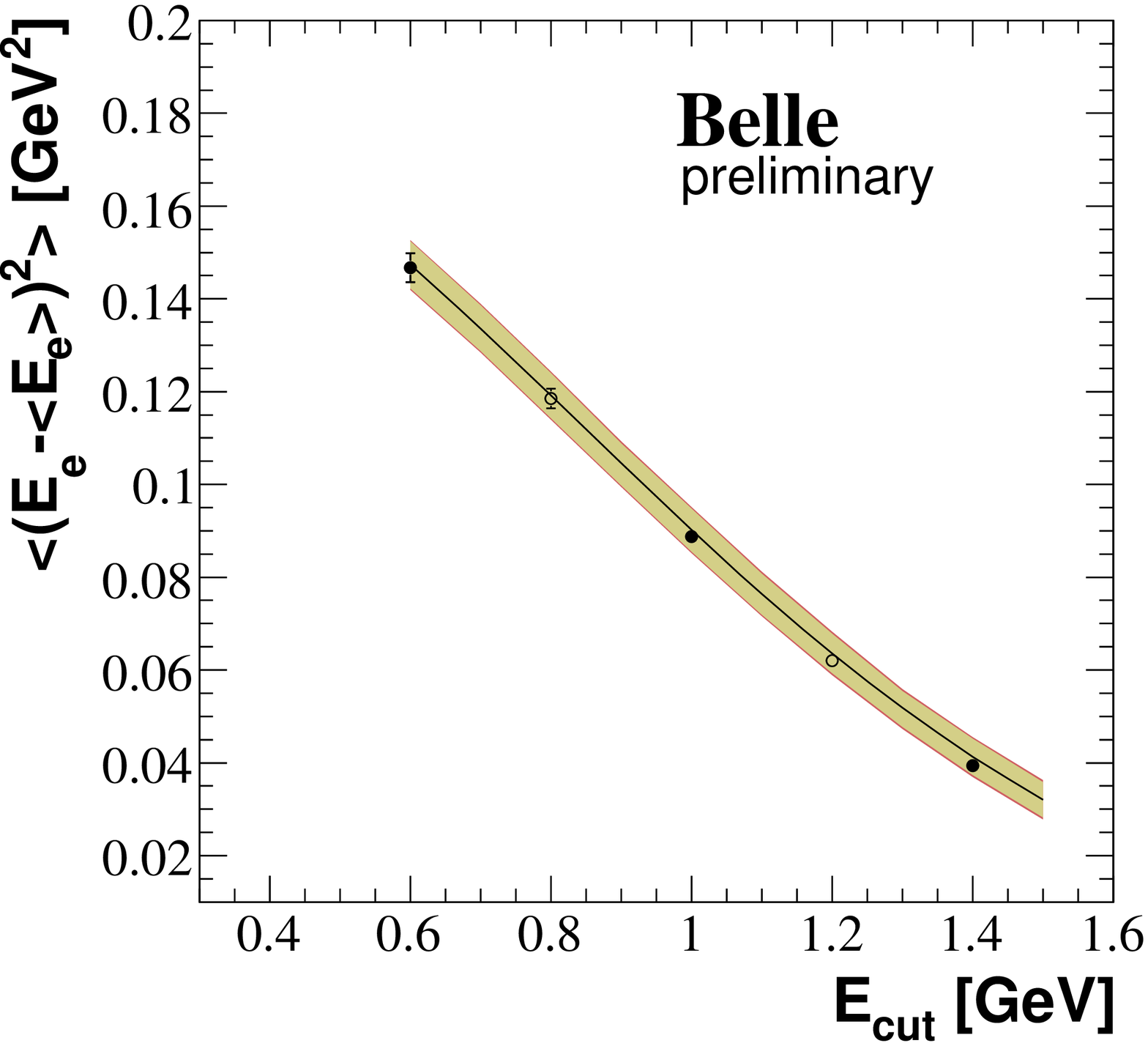}
  \includegraphics[width=0.45\textwidth]{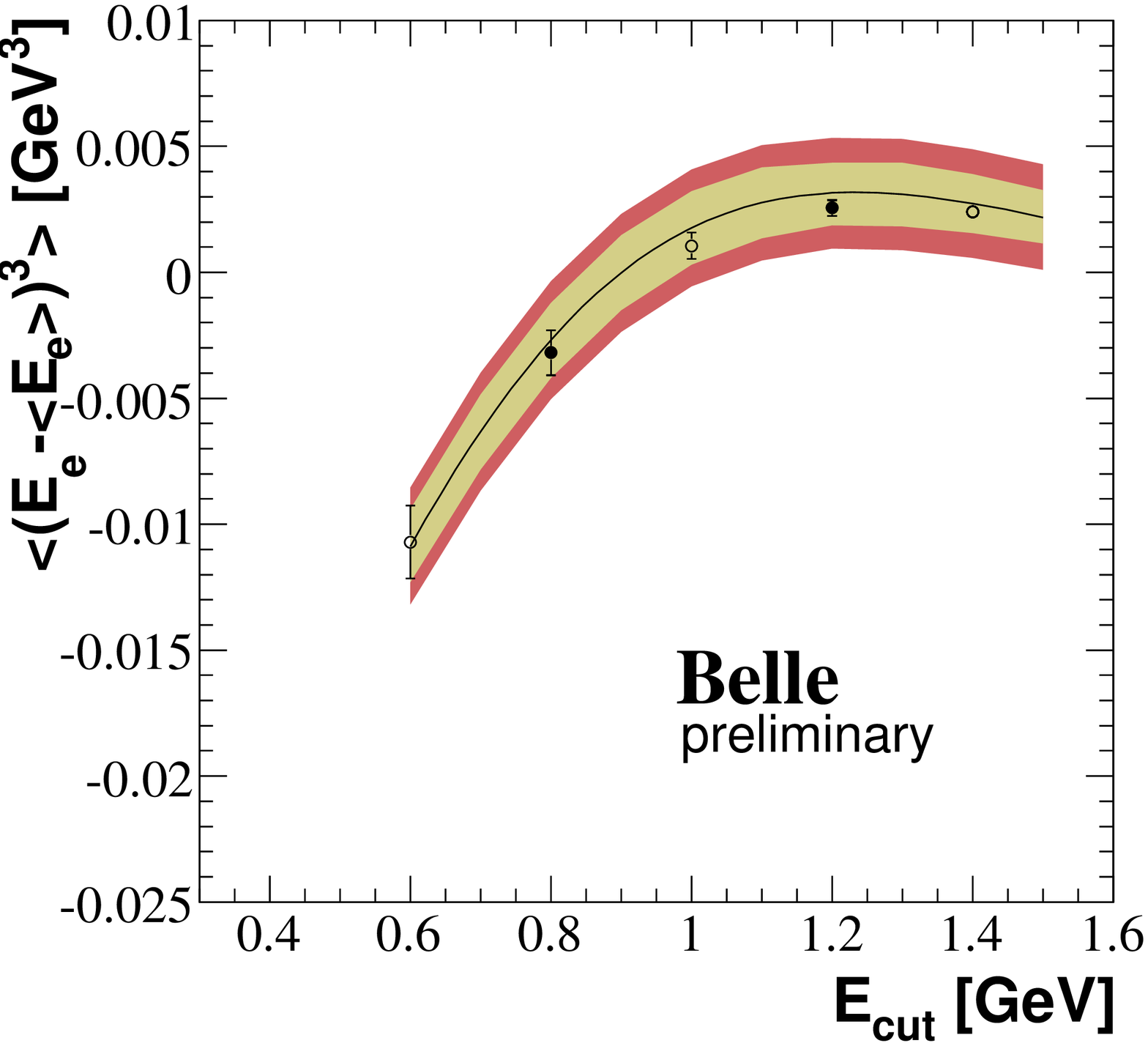}\\
  \caption{Fit results to the electron energy spectrum moments.  The
  yellow band represents the fit error, and the red band gives the
  theory and fit errors combined. Filled circles represent data points
  used in the fit, and open circles are points not used in the fit.}
  \label{fig:1}
\end{figure}

\begin{figure}
    \includegraphics[width=0.45\textwidth]{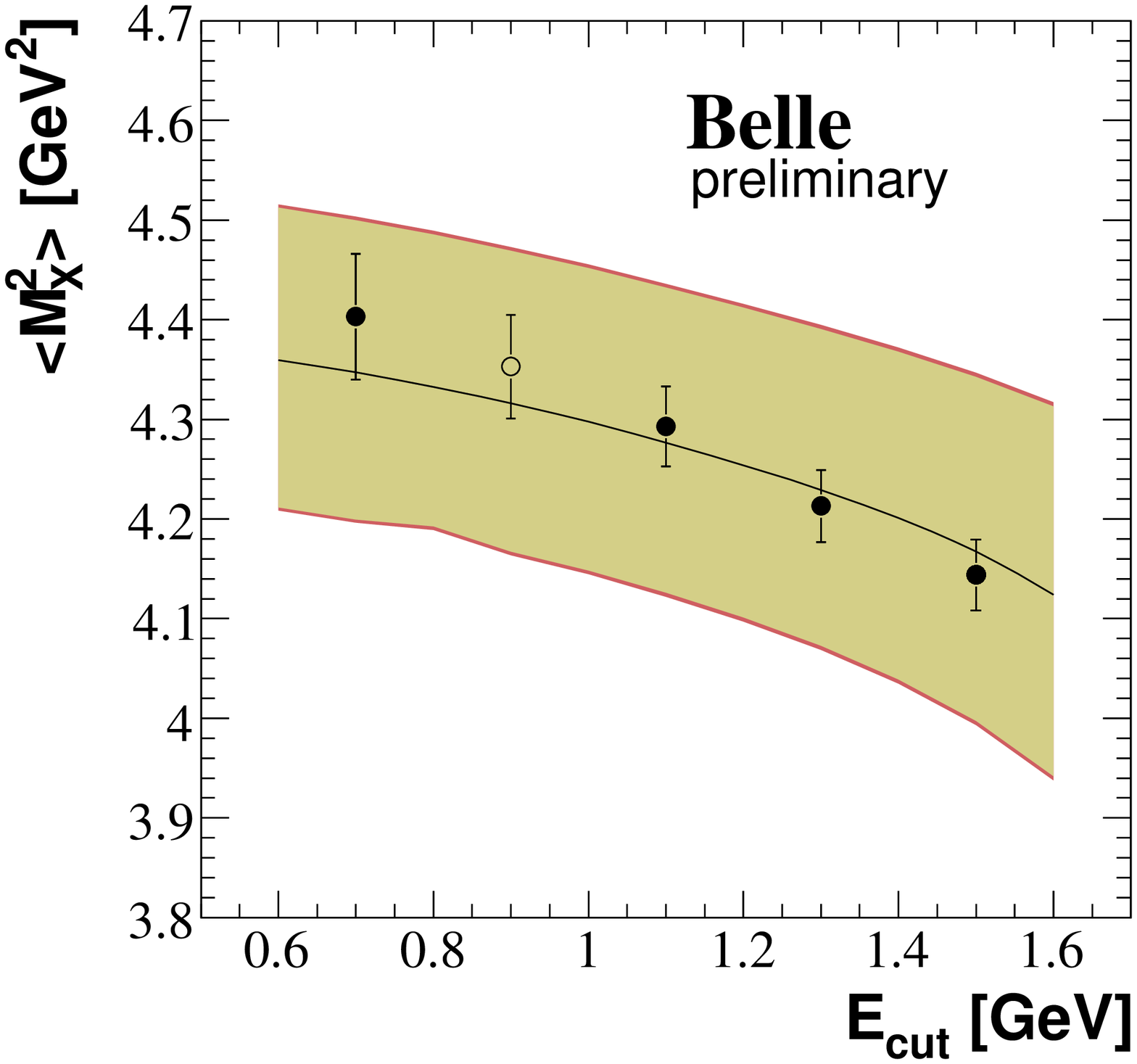}
    \includegraphics[width=0.45\textwidth]{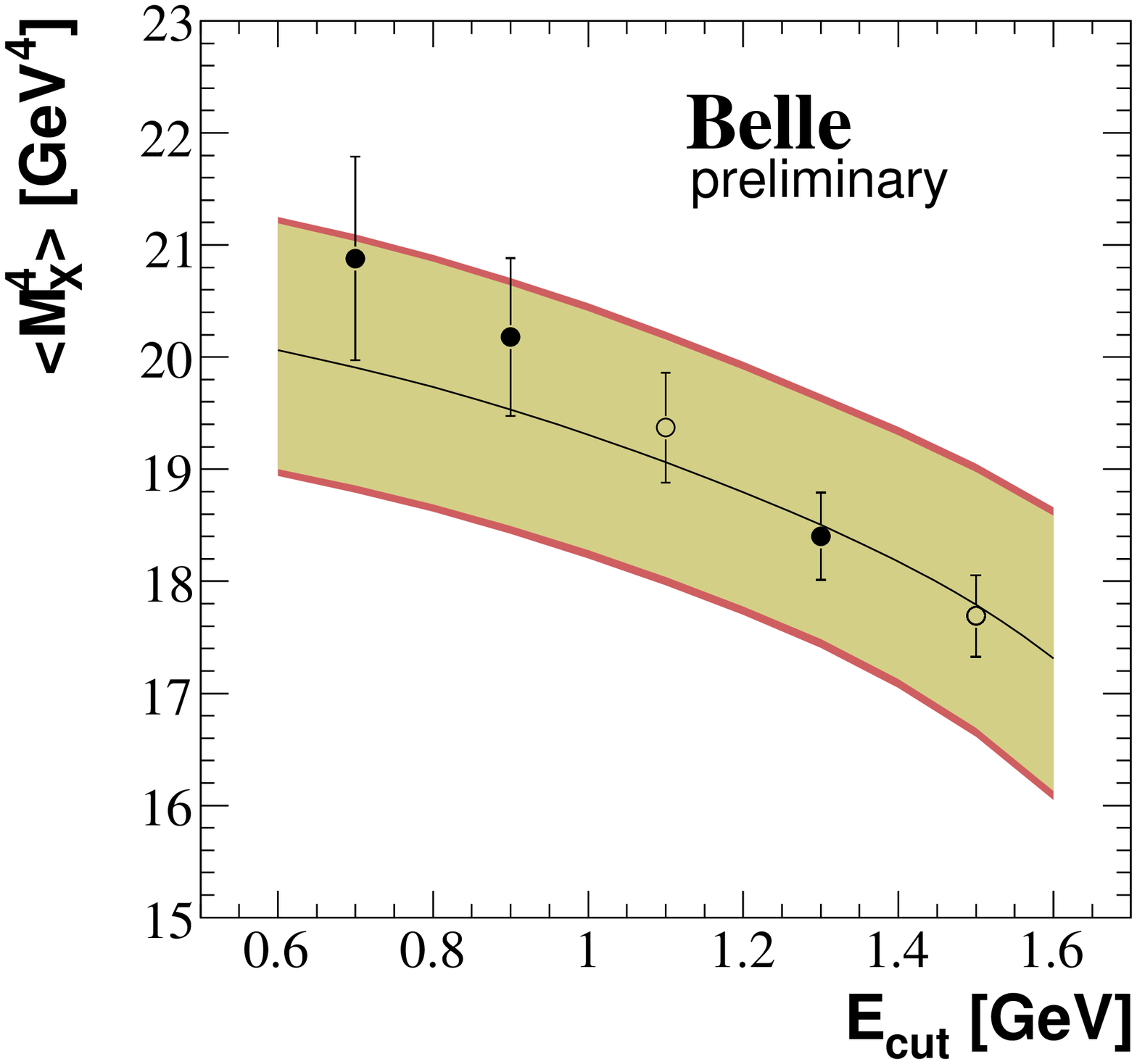}\\
    \includegraphics[width=0.45\textwidth]{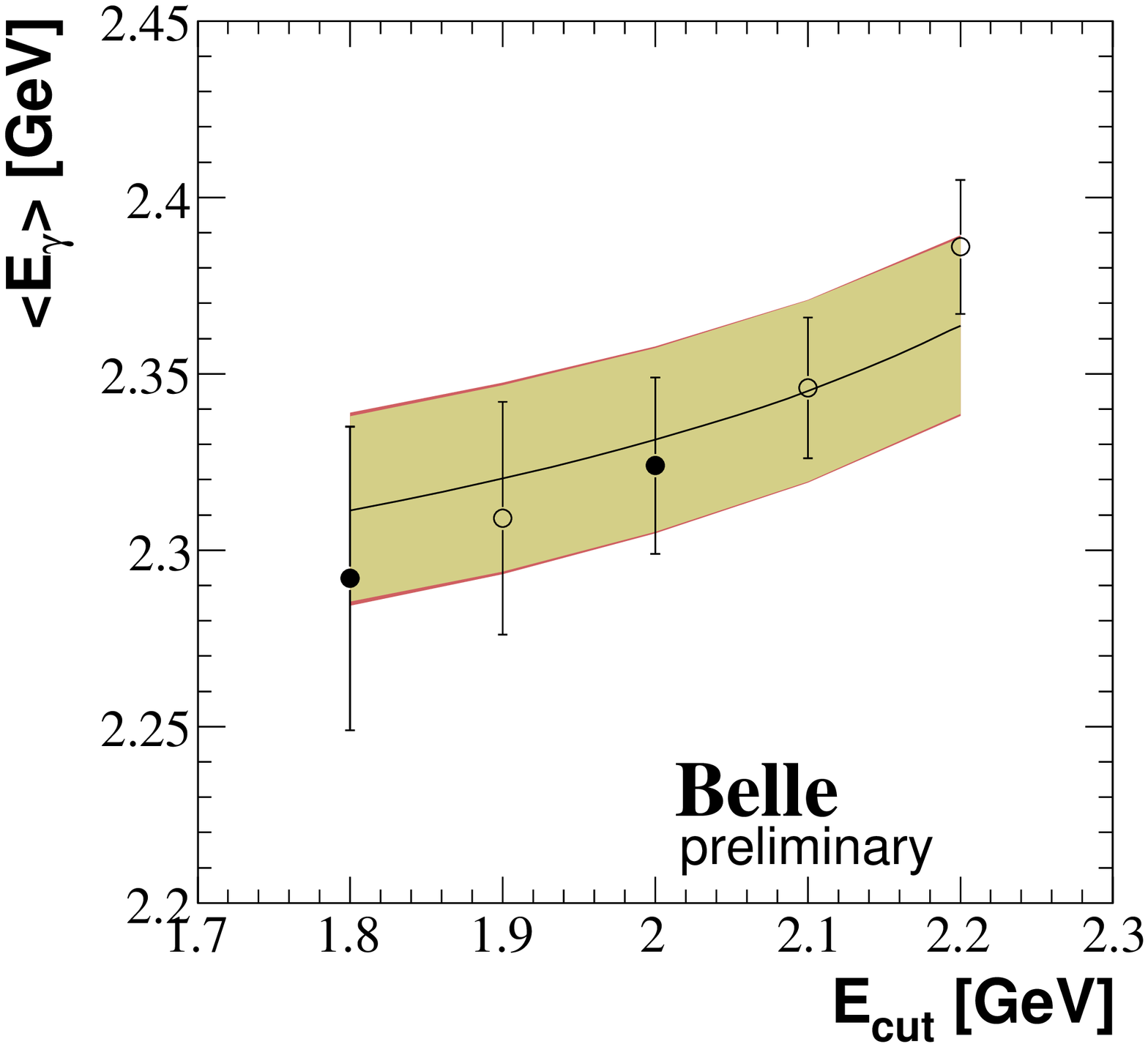}
    \includegraphics[width=0.45\textwidth]{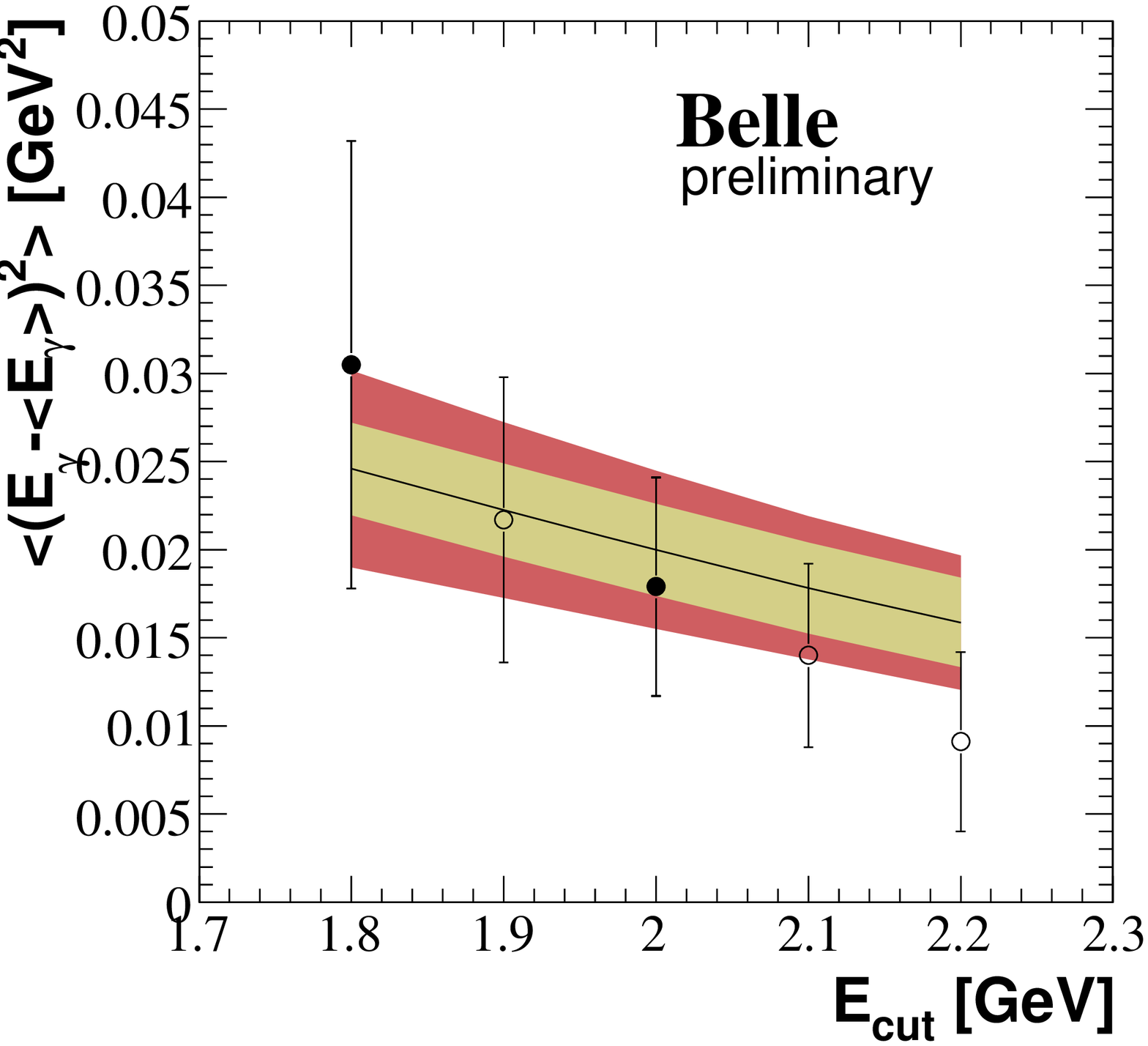}\\
    \caption{Fit results to the hadron invariant mass squared and photon
    energy spectrum moments.  The yellow band represents the fit
    error, and the red band gives the theory and fit errors
    combined. Filled circles represent data points used in the fit,
    and open circles are points not used in the fit.} \label{fig:2}
\end{figure}

Minimizing the $\chi^2$~function in Eq.~\ref{eq:2} using
MINUIT~\cite{James:1975dr}, we find the following results for the fit
parameters,
\begin{eqnarray*}
  |V_{cb}| & = & (41. 49\pm 0.52_\mathrm{fit}\pm 0.20_{\tau_B})\times
   10^{-3}~,\\
   m_b^\mathrm{1S} & = & (4.729\pm 0.048)~\mathrm{GeV}~,{\rm~and}\\
   \lambda_1 & = & (-0.30\pm 0.04)~\mathrm{GeV}^2~.
\end{eqnarray*}
The first error is the uncertainty from the fit including experimental and
theory errors, and the second error (on $|V_{cb}|$ only) is due to the
uncertainty on the average $B$~lifetime.  The correlations between
these fit parameters are provided in Table~\ref{tab:2}. Using the
measurement of the partial branching fraction at
$E_\mathrm{min}=0.6$~GeV, we obtain for the semileptonic branching
ratio (over the full lepton energy range),
\begin{equation*}
  {\mathcal B}(B\to X_c\ell\nu)=(10.62\pm 0.25)\%~.
\end{equation*}
The measured moments compared to the 1S~scheme predictions are shown
in Figs.~\ref{fig:1} and \ref{fig:2}.
\begin{table}
  \begin{center}
    \begin{tabular}{c@{\extracolsep{.1cm}}ccc}
      \hline \hline
      \rule[-1.3ex]{0pt}{4ex} & $|V_{cb}|$ & $m_b^\mathrm{1S}$ &
      $\lambda_1$\\
      \hline
      \rule{0pt}{2.7ex}$|V_{cb}|$ & 1.000& $-0.539$ & $-0.330$\\
      $m_b^\mathrm{1S}$ & & $1.000$ & $0.871$\\
      \rule[-1.3ex]{0pt}{1.3ex}$\lambda_1$ & & & 1.000\\
      \hline \hline
    \end{tabular}
  \end{center}
  \caption{Correlation coefficients of the parameters in the 1S~fit.}
  \label{tab:2}
\end{table}

\begin{figure}
  \includegraphics[width=0.45\textwidth]{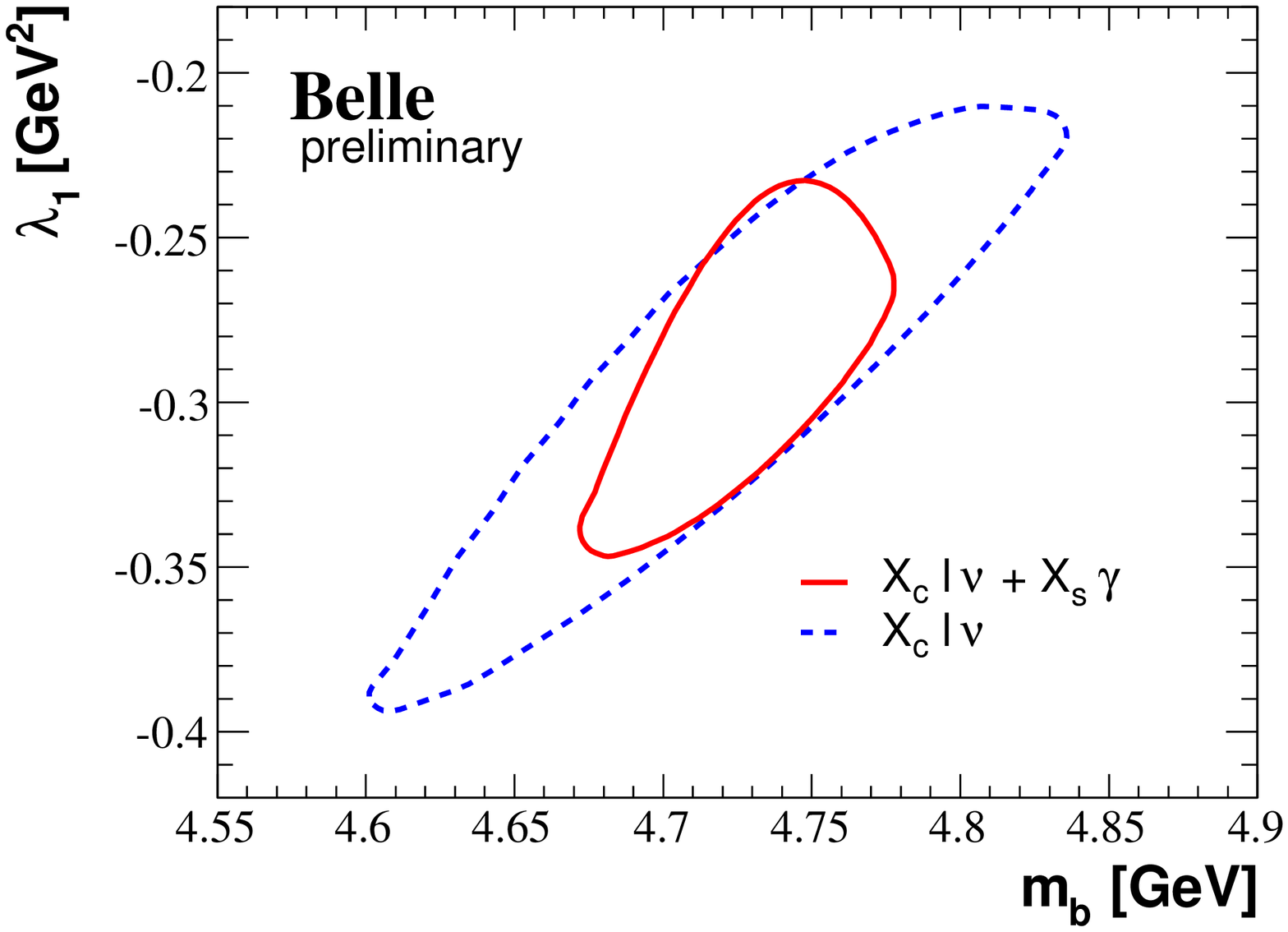}
  \includegraphics[width=0.45\textwidth]{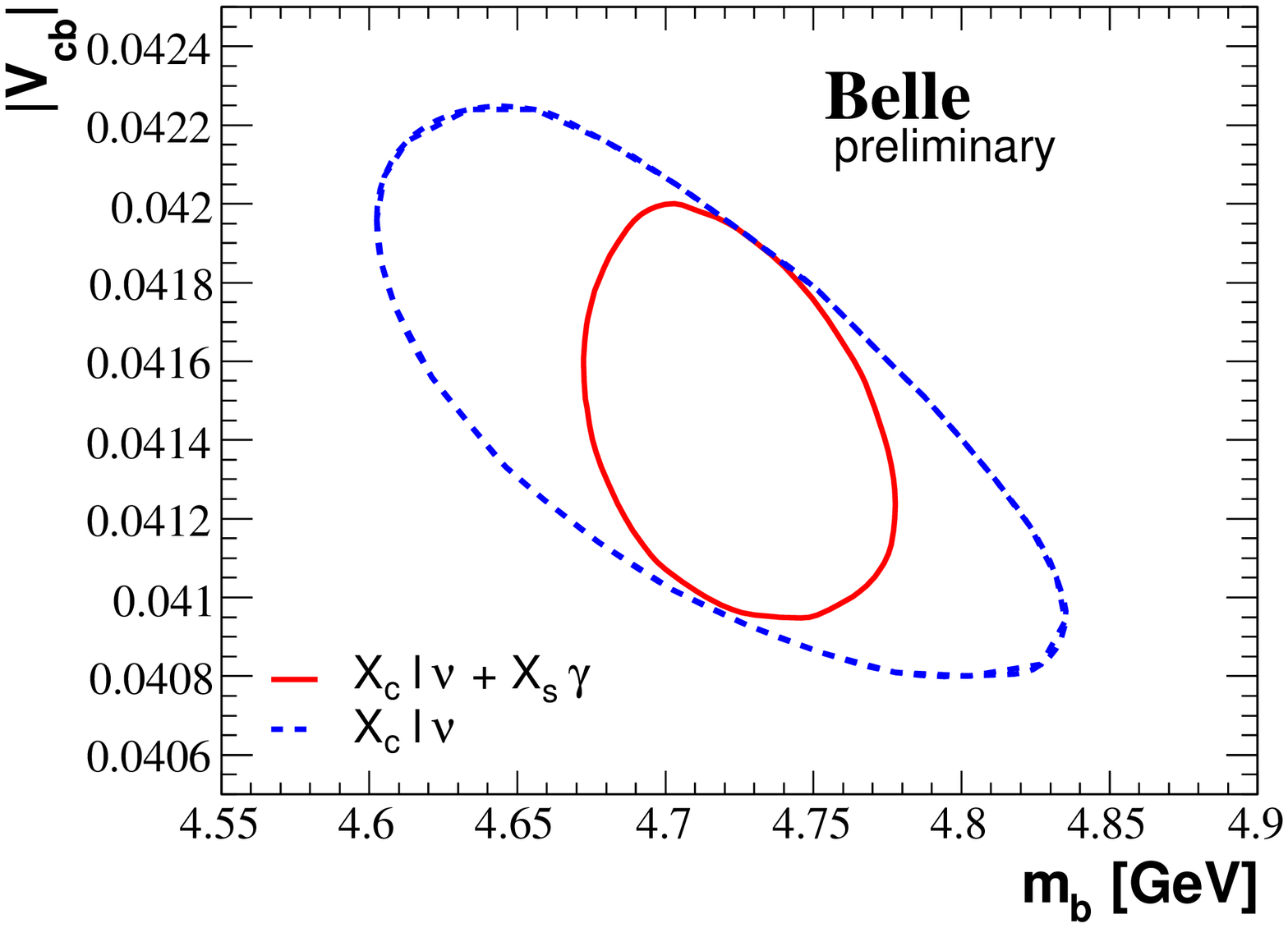}
  \caption{Fit results for $m_b^\mathrm{1S}$ and $\lambda_1$ on the left, and
    $m_b^\mathrm{1S}$ and $|V_{cb}|$ on the right. The fit to the
    $B\to X_c\ell\nu$~data only ($B\to X_c\ell\nu$ and $B\to
    X_s\gamma$~data combined) is shown by a dashed blue line (solid
    red line). The regions correspond to $\Delta\chi^2=1$.}
    \label{fig:3}
\end{figure}

We assess the stability of the fit in two ways
(Table~\ref{tab:3}): by repeating the fit only to $B\to
X_c\ell\nu$~data (21 measurements), (a) to (c); and by releasing the
$m_\chi$~constraint on the higher order parameters, (c) and (f).  To study the effect of the estimated theoretical uncertainties, we repeat the fit with all theoretical uncertainties set to zero; (b) and (e). All studies give consistent results with acceptable values of $\chi^2/\mathrm{ndf}$. Figure~\ref{fig:3} shows the contour plots for the fits
corresponding to Table~\ref{tab:3}(a) ($B\to X_c\ell\nu$~data only) and
\ref{tab:3}(d) (full fit).
\begin{table}
  \begin{center}
    \begin{tabular}{c@{\extracolsep{.1cm}}lcccccc}
      \hline \hline
      \rule[-1.3ex]{0pt}{4ex} & Data used & $m_\chi$ &
      $\sigma^2_\mathrm{theory}$ & $\chi^2/\mathrm{ndf.}$ &
      $|V_{cb}|\times 10^3$ & $m_b^\mathrm{1S}$ (GeV)& $\lambda_1$ (GeV$^2$)\\
      \hline
      \rule[-1.3ex]{0pt}{4ex}(a) & $E_{e(0,1,2,3)}$, $M^2_{X(1,2)}$ &
      0.5 & yes & 4.9/13 & $41.52\pm 0.77$ & $4.723\pm 0.103$ &
      $-0.307\pm 0.083$\\
      \hline
      \rule[-1.3ex]{0pt}{4ex}(b) & $E_{e(0,1,2,3)}$, $M^2_{X(1,2)}$ &
      0.5 & no & 9.4/13 & $41.64\pm 0.58$ & $4.655\pm 0.075$ &
      $-0.348\pm 0.052$\\
      \hline
      \rule[-1.3ex]{0pt}{4ex}(c) & $E_{e(0,1,2,3)}$, $M^2_{X(1,2)}$ &
      0.8 & yes & 1.9/13 & $42.82\pm 1.09$ & $4.588\pm 0.118$ &
      $-0.505\pm 0.271$\\
      \hline
      \rule[-1.3ex]{0pt}{4ex}(d) & $E_{e(0,1,2,3)}$,
      $E_{\gamma(1,2)}$, $M^2_{X(1,2)}$ & 0.5 & yes & 5.7/17 &
      $41.49\pm 0.52$ & $4.729\pm 0.048$ & $-0.302\pm 0.043$\\
      \hline
      \rule[-1.3ex]{0pt}{4ex}(e) & $E_{e(0,1,2,3)}$,
      $E_{\gamma(1,2)}$, $M^2_{X(1,2)}$ & 0.5 & no & 10.8/17 &
      $41.42\pm 0.45$ & $4.695\pm 0.046$ & $-0.321\pm 0.035$\\
      \hline
      \rule[-1.3ex]{0pt}{4ex}(f) & $E_{e(0,1,2,3)}$,
      $E_{\gamma(1,2)}$, $M^2_{X(1,2)}$ & 0.8 & yes & 3.9/17 &
      $42.19\pm 0.73$ & $ 4.709\pm 0.066$ & $-0.489\pm 0.087$\\
      \hline\hline
    \end{tabular}
  \end{center}
  \caption{Stability of the 1S~fit result. $\sigma^2_\mathrm{theory}$
    refers to whether or not the theory error is included in the fit.}
    \label{tab:3}
\end{table}

\section{Kinetic Scheme Analysis}

\subsection{Theoretical input}

Spectral moments of $B\to X_c\ell\nu$~decays have been derived up to
$\mathcal{O}(1/m^3_b)$ in the kinetic
scheme~\cite{Gambino:2004qm}. Compared to the original paper, the
theoretical expressions used in the fit contain an improved
calculation of the perturbative corrections to the lepton energy
moments~\cite{ref:2} and account for the $E_\mathrm{min}$~dependence
of the perturbative corrections to the hadronic mass
moments~\cite{Uraltsev:2004in}. For the $B\to X_s\gamma$~moments, the
(biased) OPE prediction and the bias correction have been
calculated~\cite{Benson:2004sg}.

All these expressions depend on the $b$- and $c$-quark masses
$m_b(\mu)$ and $m_c(\mu)$, the non-perturbative
parameters $\mu^2_\pi(\mu)$ and $\mu^2_G(\mu)$ ($\mathcal{O}(1/m^2_b)$),
$\tilde\rho^3_D(\mu)$ and $\rho^3_{LS}(\mu)$ ($\mathcal{O}(1/m^3_b)$),
and $\alpha_s$~\cite{ref:3}. The theoretical uncertainties can be
separated into two categories: non-perturbative (related to the
expansion in $1/m_b$) and perturbative (related to the expansion in
$\alpha_s$).

Following the recipe in Ref.~\cite{Gambino:2004qm}, the
non-perturbative uncertainties are evaluated by varying $\mu^2_\pi$ and
$\mu^2_G$ ($\tilde\rho^3_D$ and $\rho^3_{LS}$) by $\pm 20\%$ ($\pm 30\%$)
around their ``nominal'' values $m_b=4.6$~GeV, $m_c=1.18$~GeV,
$\mu^2_\pi=0.4$~GeV$^2$, $\tilde\rho^3_D=0.1$~GeV$^3$,
$\mu^2_G=0.35$~GeV$^2$ and $\rho^3_{LS}=-0.15$~GeV$^3$, corresponding
to the uncertainty of the respective Wilson coefficient. All these
variations are considered uncorrelated for a given moment. The
theoretical covariance matrix is then constructed by treating these
errors as fully correlated for a given moment with different
$E_\mathrm{min}$ while they are treated as uncorrelated between
moments of different order.

For the moments of the photon energy spectrum, we take 30\% of the
absolute value of the bias correction as its uncertainty. This
additional theoretical error is considered uncorrelated for moments
with different $E_\mathrm{min}$ and different order.

The theoretical uncertainties mentioned so far (non-perturbative, bias
correction) are used to construct a theoretical covariance matrix for
the measurements and are thus included in the fit. The perturbative
uncertainties are estimated by repeating the fit, setting $\alpha_s$
to a different value. For lepton and photon energy (hadronic mass)
moments, we vary $\alpha_s$ within $\pm 0.04$ ($\pm 0.1$) around the
central value 0.22 (0.3).  These ranges of variation follow the recommendations in Ref.~\cite{Gambino:2004qm}. The 
different treatment of the hadron mass moments is due to the fact 
that the calculation of the perturbative corrections to these 
moments is less complete.

\subsection{\boldmath The $\chi^2$~function}

We use a $\chi^2$~function with seven free parameters: the
semileptonic $b\to c$ branching fraction $\mathcal{B}(B\to
X_c\ell\nu)$, $m_b$, $m_c$, $\mu^2_\pi$, $\tilde\rho^3_D$, $\mu^2_G$
and $\rho^3_{LS}$,
\begin{equation}
  \chi^2 = \sum_{i,j} (\langle X\rangle_i^\mathrm{meas.}-\langle
  X\rangle_i^\mathrm{kin})\mathrm{cov}^{-1}_{ij}(\langle
  X\rangle_j^\mathrm{meas.}-\langle X\rangle_j^\mathrm{kin})~.
\end{equation}
Here, $\langle X\rangle^\mathrm{meas.}_i$ are the measured moments. $\langle X\rangle^\mathrm{kin}_i$ are the corresponding kinetic scheme predictions that depend on these free
parameters. The covariance matrix is the sum of the experimental and
theoretical error matrices.

We determine the CKM element $|V_{cb}|$ by treating it as an eighth
free parameter in the fit. $|V_{cb}|$ is related to the semileptonic
width $\Gamma(B\to X_c\ell\nu)$ by~\cite{Benson:2003kp}
\begin{eqnarray}
  \frac{|V_{cb}|}{0.0417} & = & \left(\Gamma(B\to
  X_c\ell\nu)\frac{1.55~\mathrm{ps}}{0.105}\right)^{1/2}\times
  (1-0.0018)\times (1+0.30(\alpha_s-0.22)) \label{eq10}\\
  \lefteqn{\times
  (1-0.66(m_b-4.6~\mathrm{GeV})+0.39(m_c-1.15~\mathrm{GeV})} \nonumber\\
  \lefteqn{+0.013(\mu^2_\pi-0.4~\mathrm{GeV}^2)+0.09(\tilde\rho^3_D-0.1~\mathrm{GeV}^3)}
  \nonumber \\
  \lefteqn{+0.05(\mu^2_G-0.35~\mathrm{GeV}^2)-0.01(\rho^3_{LS}+0.15~\mathrm{GeV}^3)~.}
  \nonumber
\end{eqnarray}
Using this expression, we calculate $\Gamma(B\to X_c\ell\nu)$ from
$|V_{cb}|$ and add the following term to the $\chi^2$~function,
\begin{equation}
  {\chi'}^2=\chi^2+(\frac{\mathcal{B}_{X_c\ell\nu}}{\Gamma(B\to
    X_c\ell\nu)}-\tau_B)^2/\sigma^2_{\tau_B}~.
\end{equation}

As $\mu^2_G$ and $\tilde\rho^3_{LS}$ are determined from $B^*-B$~mass
splitting and heavy quark sum rules and because the expressions depend only
weakly on these parameters, we fix $\mu^2_G$ and $\rho^3_{LS}$ to
$0.35\pm 0.07$~GeV$^2$ and $-0.15\pm 0.1$~GeV$^3$, respectively, by
adding the following terms to the $\chi^2$~function,
\begin{equation}
  {\chi''^2}={\chi'^2}+(\mu^2_G-0.35~\mathrm{GeV}^2)^2/(0.07~\mathrm{GeV}^2)^2+(\rho^3_{LS}+0.15~\mathrm{GeV}^3)^2/(0.1~\mathrm{GeV}^3)^2~.
\end{equation}
The minimization of the $\chi^{\prime \prime 2}$ is performed using
MINUIT~\cite{James:1975dr}.

\subsection{Fit results and discussion}

\begin{figure}
  \begin{center}
    \includegraphics{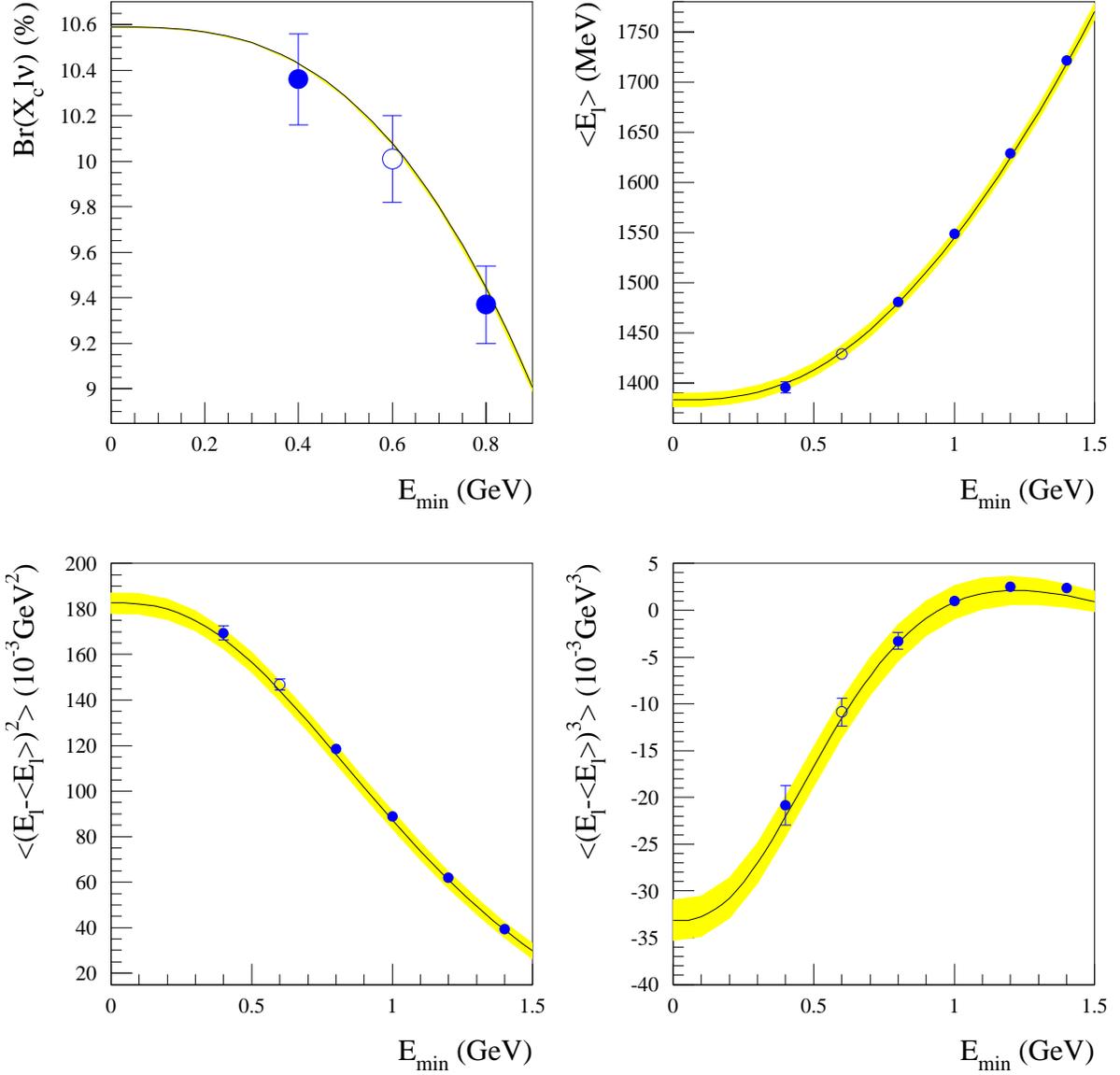}
  \end{center}
  \caption{Partial branching fractions and lepton energy moments,
    compared to the kinetic scheme fit result. The yellow bands show
    the theoretical uncertainty included in the fit (non-perturbative
    corrections, bias correction). The open symbols correspond to
    measurements not used in the fit.} \label{fig:4}
\end{figure}

\begin{figure}
  \begin{center}
    \includegraphics{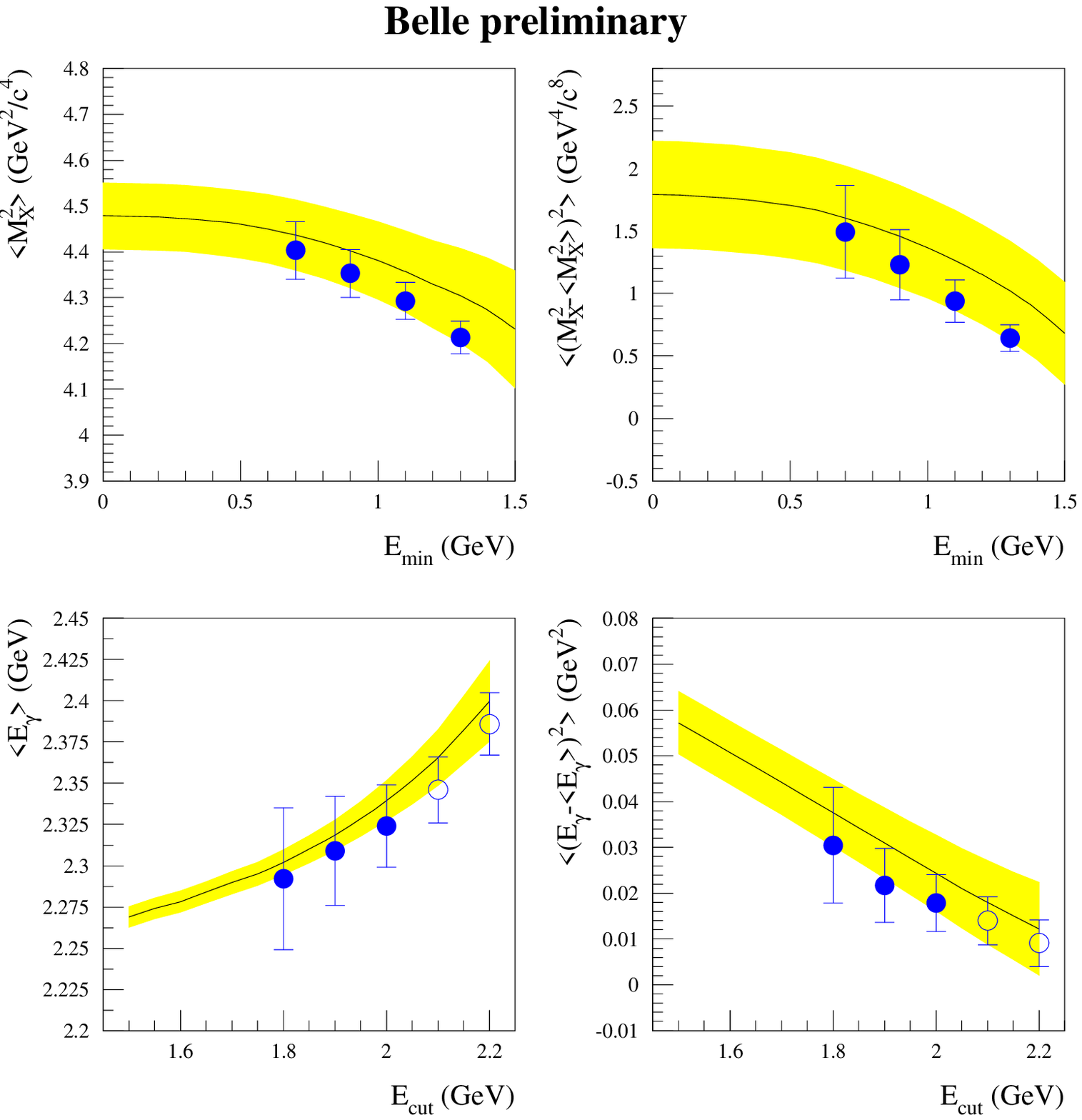}
  \end{center}
  \caption{Same as Fig.~\ref{fig:4} for the hadron mass and photon
    energy moments.} \label{fig:5}
\end{figure}

The result of the kinetic scheme analysis is shown in Table~\ref{tab:4}
and in Figs.~\ref{fig:4} and \ref{fig:5}. The value of the
$\chi^2$~function at the minimum is 17.76, compared to $(31-7)$
degrees of freedom. All results are preliminary.
\begin{table}
  \begin{center}
    {\small \begin{tabular}{c@{\extracolsep{.1cm}}ccccccc}
      \hline \hline
      \rule[-1.3ex]{0pt}{4ex} & $\mathcal{B}_{X_c\ell\nu}$ (\%) &
      $m_b$ (GeV) & $m_c$ (GeV) & $\mu^2_\pi$ (GeV$^2$) &
      $\tilde\rho^3_D$ (GeV$^3$) & $\mu^2_G$ (GeV$^2$) &
      $\rho^3_{LS}$ (GeV$^3$)\\
      \hline
      \rule{0pt}{2.7ex}value & 10.590 & 4.564 & 1.105 & 0.557 & 0.162
      & 0.358 & -0.174\\
      $\sigma$(fit) & 0.164 & 0.076 & 0.116 & 0.091 & 0.053 & 0.060 &
      0.098\\
      \rule[-1.3ex]{0pt}{1.3ex}$\sigma(\alpha_s)$ & 0.006 & 0.003 &
      0.005 & 0.013 & 0.008 & 0.003 & 0.003\\
      \hline
      \rule{0pt}{2.7ex}$\mathcal{B}_{X_c\ell\nu}$ & 1.000 & $-0.023$ &
      0.003 & 0.229 & 0.192 & $-0.147$ & $-0.024$\\
      $m_b$ & & 1.000 & 0.983 & $-0.729$ & $-0.623$ & $-0.024$ &
      $-0.111$\\
      $m_c$ & & & 1.000 & $-0.716$ & $-0.633$ & $-0.124$ & $-0.033$\\
      $\mu^2_\pi$ & & & & 1.000 & 0.851 & 0.005 & 0.052\\
      $\tilde\rho^3_D$ & & & & & 1.000 & $-0.046$ & $-0.156$\\
      $\mu^2_G$ & & & & & & 1.000 & $-0.071$\\
      \rule[-1.3ex]{0pt}{1.3ex}$\rho^3_{LS}$ & & & & & & & 1.000\\
      \hline \hline
    \end{tabular} }
  \end{center}
  \caption{Results of the kinetic scheme fit. The error from the fit
  contains the uncertainties related to the experiment, the
  non-perturbative corrections and the bias
  correction. $\sigma(\alpha_s)$ is the uncertainty related to the
  perturbative corrections. In the lower part of the table, the
  correlation matrix of the parameters is shown.} \label{tab:4}
\end{table}

\begin{figure}
  \begin{center}
    \includegraphics{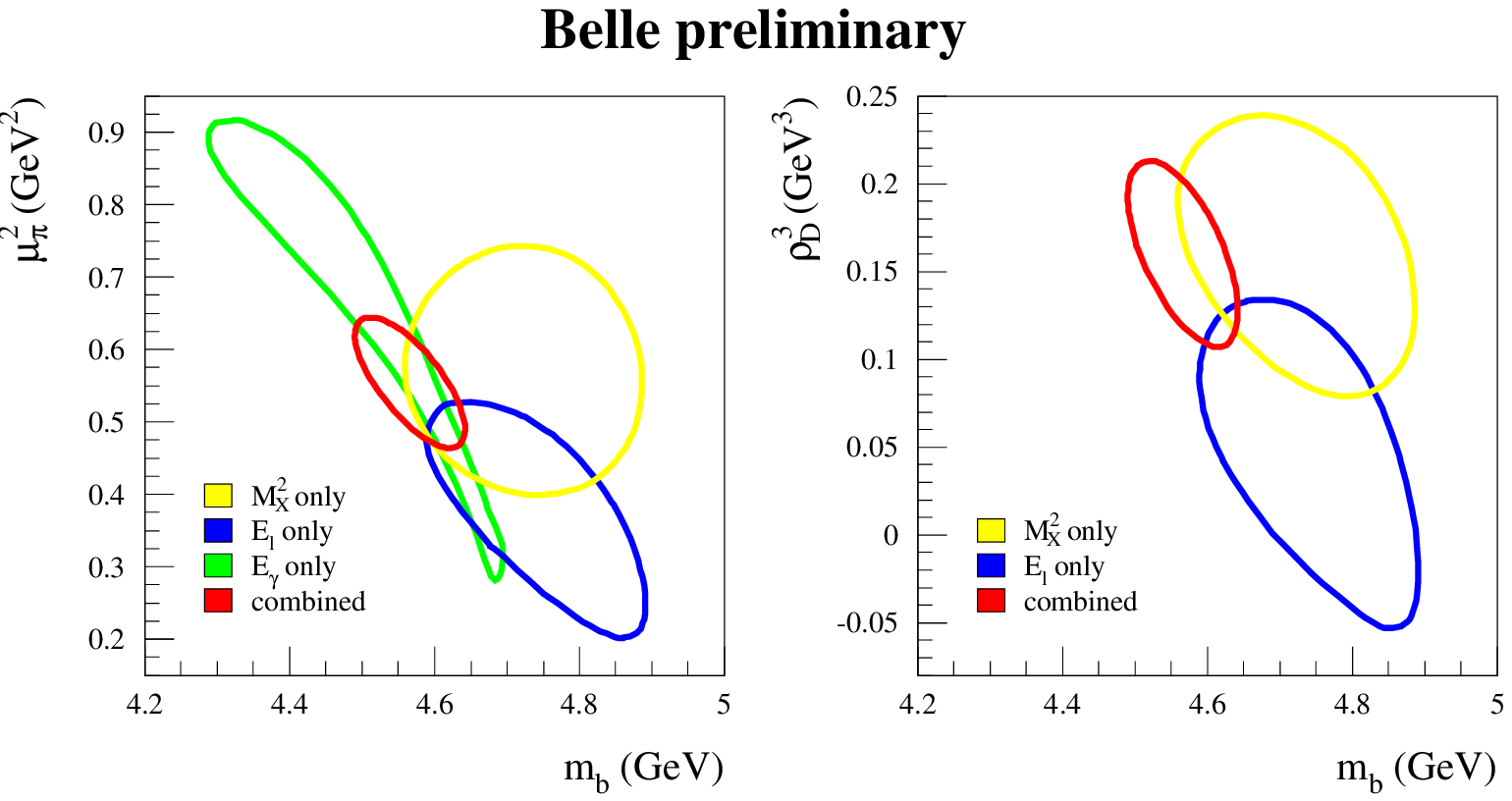}
  \end{center}
  \caption{Stability of the kinetic scheme fit. The fit is repeated
    using lepton energy moments only, hadron mass moments only and
    photon energy moments only. The ellipses are $\Delta\chi^2=1$.}
    \label{fig:6}
\end{figure}

To assess the stability of the fit, we have repeated the analysis
using lepton energy moments only, hadron mass moments only and photon
energy moments only. The result is shown in Fig.~\ref{fig:6}. In
general, changes are well covered by the fit uncertainty though the
$B\to X_s\gamma$~data seems to prefer lower values of $m_b$.

Finally, the result for $|V_{cb}|$ reads
\begin{displaymath}
  |V_{cb}|=(41.93\pm 0.65(\mathrm{fit})\pm 0.07(\alpha_s)\pm
   0.63(\mathrm{th}))\times 10^{-3}~.
\end{displaymath}
The first error is due to all uncertainties taken into account in the fit 
(experimental error in the moment measurements, non-perturbative 
corrections and bias correction to the moments, uncertainty in
$\tau_B$). The second error is obtained by varying $\alpha_s$ in the
expressions for the moments and in Eq.~\ref{eq10}. In Eq.~\ref{eq10},
we vary $\alpha_s$ by $\pm$0.008 around the central value of
0.22~\cite{ref:4}. The last error is a 1.5\% uncertainty due to the
limited accuracy of the theoretical expression for the semileptonic
width, assessed in Ref.~\cite{Benson:2003kp}.

\section{Summary}

We have performed a fit to the Belle measured spectral moments of the
lepton energy and hadronic mass spectrum in charmed semileptonic $B$
decays, and the photon energy spectrum of inclusive radiative $B$
decays using expressions for the moments in terms of HQE parameters in
the $1S$~mass and kinetic mass schemes.  The fits produce
values of $|V_{cb}|$ that are consistent between the two schemes.  In the 1S
scheme analysis we find $|V_{cb}|=(41.49\pm 0.52{\rm (fit)}\pm
0.20(\tau_B))\times 10^{-3}$, and in the kinetic scheme we obtain
$|V_{cb}|=(41.93\pm 0.65{\rm (fit)}\pm 0.07{\rm (\alpha_s)}\pm
0.63{\rm (th)})\times 10^{-3}$. The heavy quark parameters,
$m_b^{\mathrm{kin},1S}$ and $\lambda_{1},~\mu_{\pi}^2$, have been
extracted with values that are consistent with previous
determinations~\cite{Bauer:2004ve,Buchmuller:2005zv}.  Constant
feedback between theory and experiment should further confirm the
understanding of the OPE in all measurable regions of phase space.
The accuracy achieved by the Belle measurements is unprecedented by
any other experiment and the uncertainty on the heavy quark parameters
and $|V_{cb}|$ reflect these improvements.

\section{Acknowledgments}

We thank the theorists working on the $1S$~scheme: C.W.~Bauer,
Z.~Ligeti, M.~Luke, A.V.~Manohar and M.~Trott, and those working on
the kinetic scheme: P.~Gambino, N.~Uraltsev and I.~Bigi for providing
the Mathematica and Fortran code that describe the respective
calculations.
We thank the KEKB group for the excellent operation of the
accelerator, the KEK cryogenics group for the efficient
operation of the solenoid, and the KEK computer group and
the National Institute of Informatics for valuable computing
and Super-SINET network support. We acknowledge support from
the Ministry of Education, Culture, Sports, Science, and
Technology of Japan and the Japan Society for the Promotion
of Science; the Australian Research Council and the
Australian Department of Education, Science and Training;
the National Science Foundation of China and the Knowledge
Innovation Program of the Chinese Academy of Sciences under
contract No.~10575109 and IHEP-U-503; the Department of
Science and Technology of India; 
the BK21 program of the Ministry of Education of Korea, 
the CHEP SRC program and Basic Research program 
(grant No.~R01-2005-000-10089-0) of the Korea Science and
Engineering Foundation, and the Pure Basic Research Group 
program of the Korea Research Foundation; 
the Polish State Committee for Scientific Research; 
the Ministry of Science and Technology of the Russian
Federation; the Slovenian Research Agency;  the Swiss
National Science Foundation; the National Science Council
and the Ministry of Education of Taiwan; and the U.S.\
Department of Energy.


\begin{thebibliography}{99}
\bibitem{ref:1}
  The measurements of $<M_X^4>$ and $<(M_X^2 - <M_X^2>)^2>$ correspond
  to the same order in $M^2_X$ and are counted only once.
\bibitem{ref:3}
  In this analysis, all non-perturbative parameters in the kinetic
  scheme are defined at the scale $\mu=1$~GeV.
\end{thebibliography}
\end{document}